\documentclass{elsart3}

\usepackage{times}

\usepackage{graphicx}

\usepackage{epsfig,citesort,amsmath,amssymb}

\usepackage{amssymb}
\begin{document}

\begin{frontmatter}


\title{
Multicanonical Simulations of the Tails of the Order-Parameter 
Distribution of the Two-Dimensional Ising Model}
%
%
\author[ia1,ia2]{Rudolf Hilfer},
\author[ia1,ia3]{Bibhu Biswal},
\author[ia4]{Hans-Georg Mattutis} and
\author[ia5]{Wolfhard Janke\corauthref{cor1}}
\corauth[cor1]{Corresponding Author:}
\ead{wolfhard.janke@itp.uni-leipzig.de}
\address[ia1]{ICA-1, Universit{\"a}t Stuttgart,
Pfaffenwaldring 27, 70569 Stuttgart, Germany}
\address[ia2]{Institut f{\"u}r Physik, Universit{\"a}t Mainz,
55099 Mainz, Germany}
\address[ia3]{Department of Physics, Sri Venkateswara College,
University of Delhi,\\ New Delhi -- 110~021, India}
\address[ia4]{Tokyo University of Electro-communications,
Dept. of Mechanical and Control Engineering, Chofu,
Tokyo 182-8585, Japan}
\address[ia5]{Institut f{\"u}r Theoretische Physik,
Universit{\"a}t Leipzig, Augustusplatz 10/11,
04109 Leipzig, Germany}
%

\begin{abstract}
We report multicanonical Monte Carlo simulations of
the tails of the order-parameter distribution of the
two-dimensional Ising model for fixed boundary
conditions. Clear numerical evidence for ``fat''
stretched exponential tails is found below the critical temperature,
indicating the possible presence of fat tails at the critical
temperature. 
\end{abstract}

\end{frontmatter}

\vspace*{-0.3cm}
\section{Introduction}
\vspace*{-0.1cm}

\label{sec1}
A quantity of central importance for finite-size scaling (FSS) analyses
of critical phenomena is the order-parameter distribution $p(m)$ 
\cite{car88}.
Most properties of $p(m)$ at criticality
are known from computer simulations
\cite{bin81}. Analytical information comes from
field theoretic renormalization group calculations
\cite{BZ85}, conformal field theory \cite{BD85}, and
also a generalized classification theory of phase transitions
~\cite{hil91f,hil94d}.
While some of the analytical predictions seem to have
been corroborated by numerical simulations ~\cite{hil94d,hil94e},
the predictions for the tails of the critical order-parameter
distribution could not be confirmed. 
Recording of the very small probabilities in the tails
requires special techniques such as multicanonical 
simulations \cite{bn92}. 
Many recent studies \cite{hil94e,SB95b} have
attempted this, but failed in establishing the true behaviour of the
tails of $p(m)$. 

\begin{figure*}[tbh]
\vspace*{0.05cm}
\begin{center}
\begin{minipage}{15cm}
\epsfig{angle=-90,width=7.32cm,figure=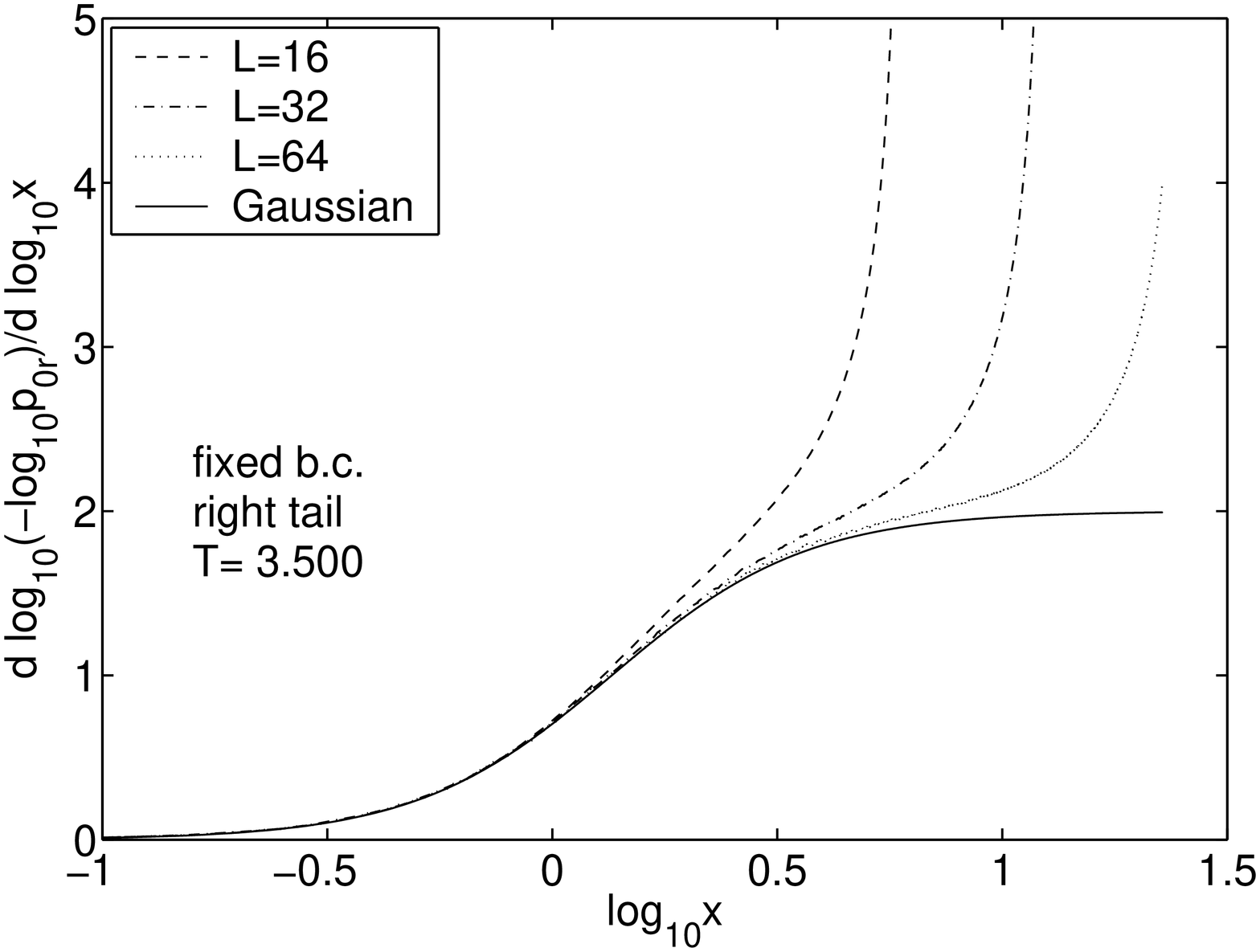} \hspace*{0.5cm}
\epsfig{angle=-90,width=6.90cm,figure=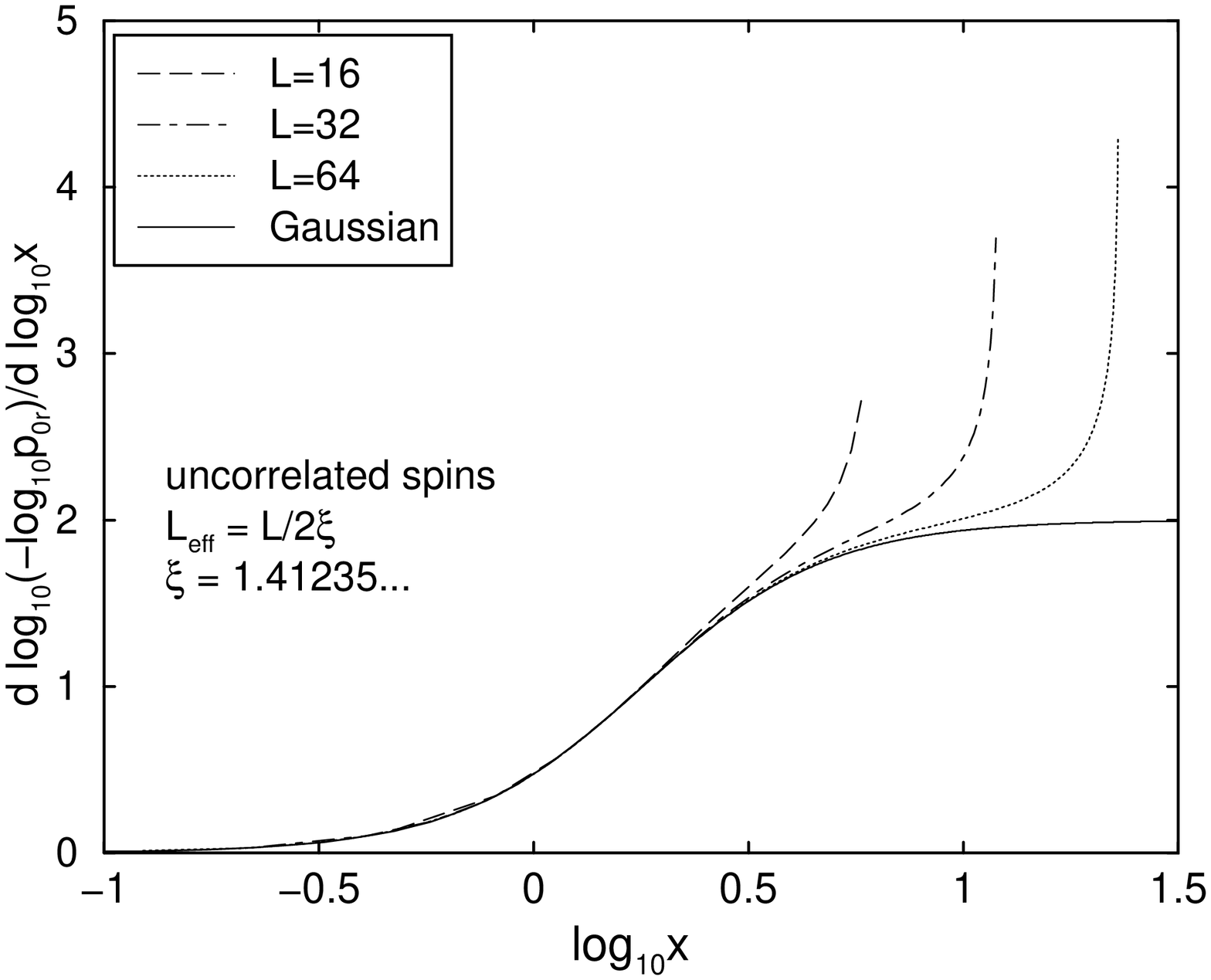}\\
\epsfig{angle=-90,width=7.32cm,figure=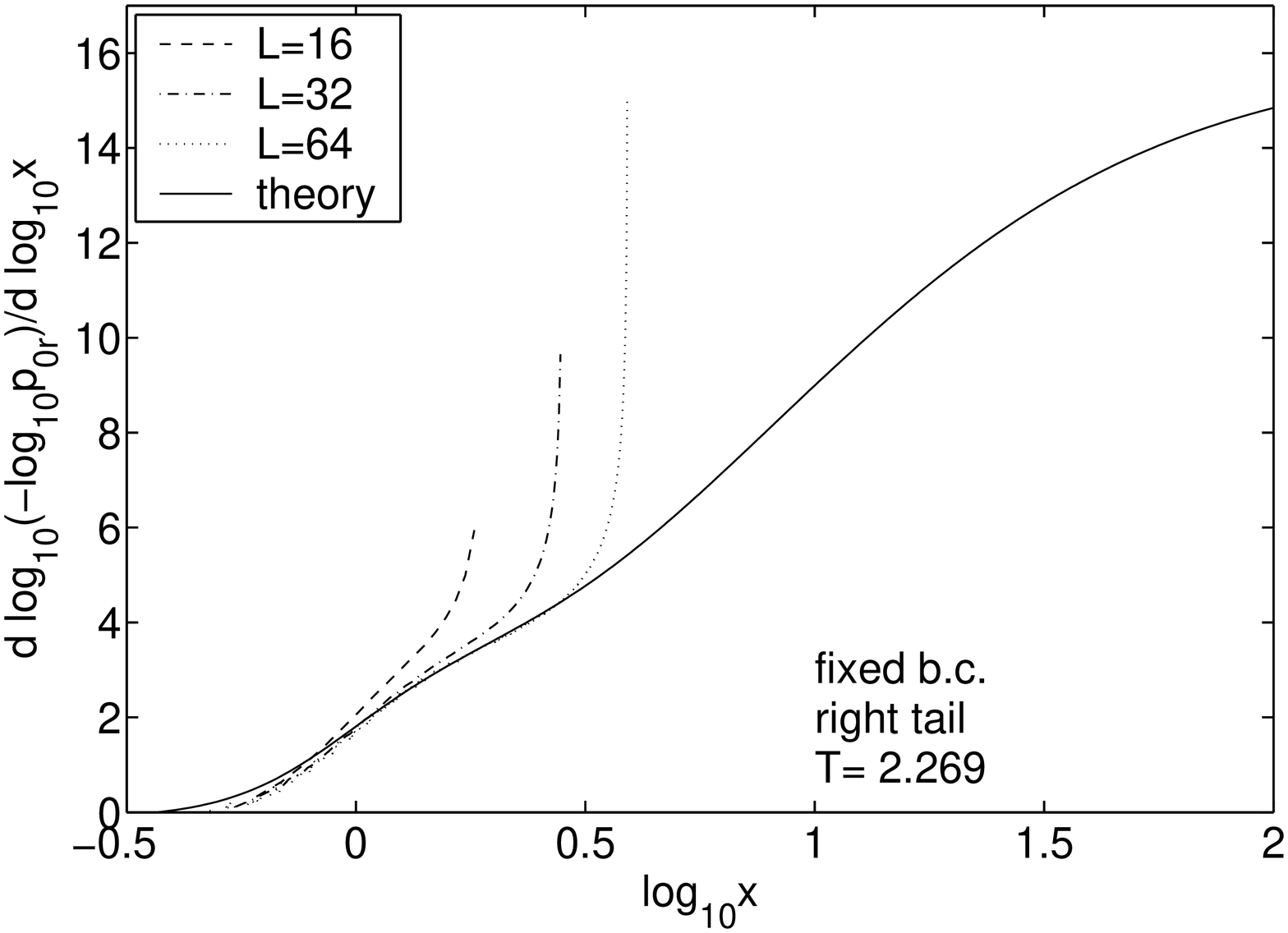} \hspace*{0.5cm}
\epsfig{angle=-90,width=6.90cm,figure=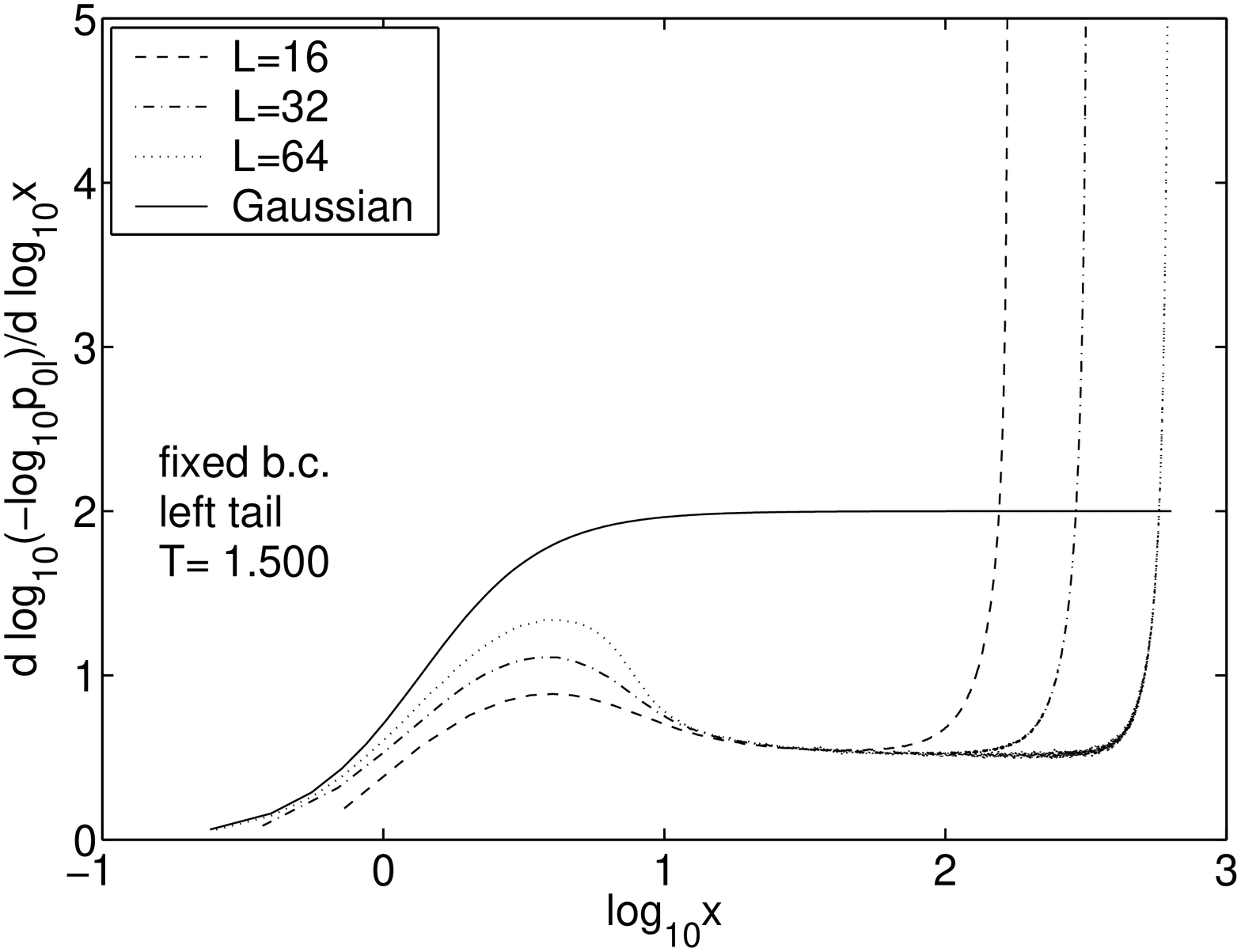} 

\caption{\label{fig:high} Tail analyses for the 2D Ising model
with fixed boundary conditions.
}
\end{minipage}
\end{center}
\end{figure*}

In the present paper we report results of high precision
multicanonical Monte Carlo simulations for the two-dimensional (2D) 
Ising model on square lattices with fixed (i.e. all boundary spins fixed
to $+1$) boundary conditions. One of the objectives is to study
whether the order-parameter distribution obtained from the simulation
can be considered to be asymptotic with respect
to system size.
A secondary objective is to study fixed (symmetry breaking) boundary
conditions because the asymmetry 
should give rise to an asymmetry in the far tail 
behaviour. 

\section{Analysis methods}

We consider the 2D Ising model on a square lattice 
with $N=L^2$ spins $\sigma_i=\pm 1$ interacting
according to the usual 
Hamiltonian
\begin{eqnarray}
{H}=-J\sum_{\langle ij \rangle} \sigma_i \sigma_j,
\label{eq:H}
\end{eqnarray}
where $J>0$ is the ferromagnetic coupling strength
and the summation $\sum_{\langle ij \rangle}$ runs over all nearest neighbour 
pairs on the lattice.  
The order parameter is the magnetization per spin,
$-1 \leq m = (1/N) \sum_{i=1}^N \sigma_i \leq 1$. 
In the following we set $J=1$,
the Boltzmann constant to unity and denote the temperature by
$T$.

The probability density $p(m)$ of the
order parameter depends parametrically on temperature and
system size, $p(m) = p(m;T,L)$. At criticality, traditional FSS predicts 
the scaling form \cite{car88}
\begin{eqnarray}
p(m;T=T_c,L) = L^{\beta/\nu} \tilde{p}(mL^{\beta/\nu}),
\end{eqnarray}
where $\tilde{p}(\tilde{m})$ with $\tilde{m} = m L^{\beta/\nu}$ 
is a universal scaling 
function. Since for the 2D Ising model
$\nu = 1$ and $\beta = 1/8$, the scaling variable is thus 
$\tilde{m} = m L^{1/8}$.
For the analysis of the tails of $p(m)$ we first determine constants $A, B, C$
such that $p_0(x) = A p(B(m-C))$ has mean zero, unit norm and unit variance.
Then we split the peak into its
left and right tails by defining the functions
\begin{eqnarray}
p_{0l}(x) &=& p_0(x_{\rm peak} - x) \qquad \text{for $x < x_{\rm peak}$} \, , \\
p_{0r}(x) &=& p_0(x - x_{\rm peak}) \qquad \text{for $x > x_{\rm peak}$} \, ,
\end{eqnarray}
where $x_{\rm peak}$ is the position of the maximum.
To exhibit stretched
exponential tails we finally calculate the logarithmic derivatives
\begin{eqnarray}
q(y) &=& \frac{\d\log_{10}(-\log_{10}p_{0i})}{\d\log_{10}x} \, ,
\end{eqnarray}
where $i=l,r$, and plot them against $y=\log_{10}x$. In this
representation, tails of the form 
$p_0(x) \sim B(x+c)^\beta \exp[-A(x+c)^\alpha]$ 
lead to a plateau at the value $\alpha$ for large $x$. A
standard Gaussian distribution $(1/\sqrt{2\pi})\exp(-x^2/2)$
thus approaches asymptotically the value 2.

\section{Results}

We performed multicanonical simulations \cite{bn92} by
flattening the probability distribution of the magnetization
for system sizes up to $L=64$ at three characteristic 
temperatures.
The right tails in the high-temperature
phase at $T=3.5$ are shown in the upper left plot of 
Fig.~1. With increasing system size the distribution
approaches Gaussian behaviour as is expected from the
central limit theorem for a finite correlation
length $\xi = 1/[-\ln \tanh(1/T)- 2/T] = 1.41235\dots$.
In fact, assuming heuristically that there are effectively
$L_{\rm eff}^2$ uncorrelated spins with $L_{\rm eff} =
L/2\xi$ ($ = 6, 12, 23$ for $L = 16, 32, 64$), we obtain 
the very similar looking curves in the upper right plot of Fig.~1,
explaining the finite-size effects.

At the critical temperature $T_c = 2/{\rm arcsinh}(1) = 2.2691\dots$,
theory predicts a right tail of the form 
$x^{(\delta-1)/2} \exp(-x^{\delta+1})$,
where
$\delta = 15$ is the equation of state exponent, leading to a
plateau at 16 \cite{hil94e}. 
Our data shown in the lower left plot of Fig.~1 reveal a shoulder 
developing with increasing $L$. 
The convergence, however, is 
extremely
slow \cite{ours}.
The solid line is a
fit using this theoretical prediction,
\begin{eqnarray}
p_0(x)=a((x+c)/b)^7\exp(-((x+c)/b)^{16}),
\label{eq:theory-fit}
\end{eqnarray}
with $a=1.31$, $b=8.59$, and $c=7.79$. 

\begin{figure}
\epsfig{width=7.32cm,figure=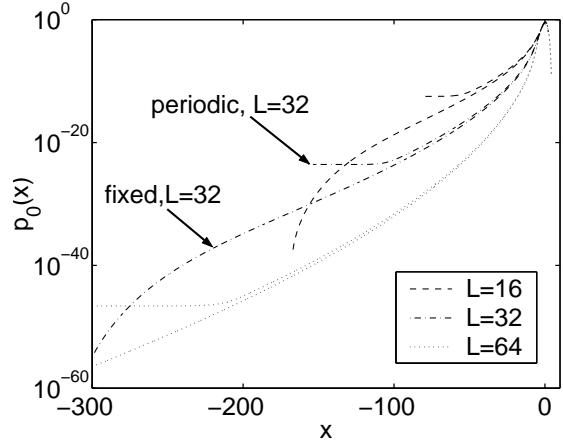}
\caption{Rescaled order parameter distributions
$p_0(x)$ on a logarithmic scale for $T=1.5$ and 
$L = 16, 32, 64$ for fixed and periodic boundary conditions.
For periodic boundary conditions only the right half of the 
distribution is shown.}
\label{fig:rescaled_lowT}
\end{figure}

For low temperatures
the magnetization is close to unity. At $T=1.5$ a true peak
starts developing only for system sizes $L > 16$, and even for $L=32$ only 
five data points exist to the right of the peak, cf.\ Fig.~\ref{fig:rescaled_lowT}. 
We therefore found only
weak evidence for Gaussian behaviour 
with increasing $L$ for the right tails \cite{ours}. 
Here the left tails shown in the lower right plot of 
Fig.~1 are more interesting. 
Near the peak (i.e. for
small $x$) we see a narrow regime over which
the curves approach Gaussian behaviour as one would expect
again from the central limit theorem.
In the intermediate range of $x$, the curves show a
plateau at $0.5$, corresponding to a fat stretched exponential
$\sim\exp(-\sqrt{x})$. This represents the well known droplet regime
predicted analytically \cite{shl89}. We observed this stretched
exponential tail 
in the distributions 
for low temperatures all the way up to $T_c$.
For periodic boundary conditions the 
stretched exponential 
tails eventually cross over into a flat bottom, 
reflecting phase coexistence on finite
lattices governed by strip-like spin configurations.
For fixed boundary conditions the
same stretched exponential tail is terminated by a cutoff function. 
Visually the three different regimes can also easily be identified
in the logarithmic plot of $p_0(x)$ shown in Fig.~\ref{fig:rescaled_lowT}.

\section{Summary}

To summarize, 
the order-parameter distribution 
at the critical point
is found to approach its asymptotic universal scaling function extremely 
slowly. This is also true for periodic boundary conditions \cite{ours}.
Our extrapolations discussed in detail in Ref.~\cite{ours} indicate that
the required system sizes ($L \gtrsim 10^5$) are beyond present day
numerical resources even for the 2D Ising model.


\begin{thebibliography}{00}

\bibitem{car88}
%
J. Cardy (ed.), {\em Finite-Size Scaling\/}, 
North-Holland, Amsterdam, 1988; V. Privman (ed.), 
{\em Finite-Size Scaling and Numerical Simulation of
  Statistical Systems\/}, World Scientific, Singapore, 1990.

\bibitem{bin81}
K. Binder, Z. Phys. B 43 (1981) 119;
A. Bruce, J. Phys. C 14 (1981) 3667;
J. Kim, A. Souza, D.P. Landau, Phys. Rev. E 54 (1996) 2291;
H. Bl{\"o}te, J. Heringa, M.Tsypin, Phys. Rev. E 62 (2000) 77.

\bibitem{BZ85}
E. Brezin, J. Zinn-Justin, Nucl. Phys. B 257 (1985) 867;
A. Esser, V. Dohm, M. Hermes, J. Wang, Z. Phys. B 97 (1995) 205;
X. Chen, V. Dohm, A. Talapov, Physica A 232 (1996) 375.

\bibitem{BD85}
T. Burkhardt, B. Derrida, Phys. Rev. B 32 (1985) 7273.

\bibitem{hil91f}
R. Hilfer, Physica Scripta 44 (1991) 321;
Phys. Rev. Lett. 68 (1992) 190;
Mod. Phys. Lett. B 6 (1992) 773;
Int. J. Mod. Phys. B 7 (1993) 4371;
Phys. Rev. E 48 (1993) 2466.

\bibitem{hil94d}
R. Hilfer, Z. Physik B 96 (1994) 63.

\bibitem{hil94e}
R. Hilfer, N.B. Wilding, J. Phys. A: Math. Gen. 28 (1995) L281.

\bibitem{bn92}
B.A. Berg, T. Neuhaus, Phys. Lett. B 267 (1991) 249; 
Phys. Rev. Lett. 68 (1992) 9; 
W. Janke, Int. J. Mod. Phys. C 3 (1992) 1137; 
Physica A 254 (1998) 164.

\bibitem{SB95b}
G. Smith, A. Bruce, J. Phys. A: Math. Gen. 28 (1995) 6623;
D. Stauffer, Int. J. Mod. Phys. C 9 (1998) 625;
M. Tsypin, H. Bl{\"o}te, Phys. Rev. E 62 (2000) 73.

\bibitem{ours}
R. Hilfer, B. Biswal, H.G. Mattutis, W. Janke, Phys. Rev. E 68
(2003) 046123.

\bibitem{shl89}
S.B. Shlosman, Comm. Math. Phys. 125 (1989) 81.

\end{thebibliography}
\end{document}